\documentclass[aps,twocolumn]{revtex4}
\usepackage{epsf}
\newcommand{\Av}{\mbox{{\bf A}}}
\newcommand{\Rv}{\mbox{{\bf R}}}

\newcommand{\lv}{\mbox{{\bf l}}}
\begin{document}
\title{Vortex structure in {\it d}--density wave scenario of pseudogap}
\author{Maciej M. Ma\'ska and Marcin Mierzejewski} 
\email{marcin@server.phys.us.edu.pl}
\affiliation{
Department of Theoretical Physics, Institute of Physics, 
University of Silesia, 40-007 Katowice,
Poland}
\begin{abstract}
We investigate the vortex structure assuming the {\it d}--density 
wave scenario of the pseudogap. 
We discuss the 
profiles of the order parameters in the vicinity of the vortex, 
effective vortex charge and the local density of states. 
We find a pronounced modification of these quantities when
compared to a purely superconducting case. Results have been obtained
for a clean system as well as in the presence of a nonmagnetic impurity.     
We show that the competition  
between superconductivity and the density wave may explain some
experimental data recently 
obtained for high--temperature superconductors. In particular, we show that
the {\it d}--density wave scenario explains the asymmetry of the gap
observed in the vicinity of the vortex core.
\end{abstract}
\pacs 74.25.Ha 74.25.Op 74.20.-z 
\maketitle
\section{Introduction}
Recent scanning--tunneling--microscopy (STM) experiments give a new 
insight into the electronic structure of the vortex cores in the 
high--temperature superconductors (HTSC), and, more generally, into 
the problem of the interplay between superconductivity and a magnetic 
field. In type--II superconductors, the flux penetrating the vortex 
suppresses the superconducting order parameter locally in the core. 
It was shown in 1964 by Caroli {\it at al.}\cite{caroli} that there should 
be bound states around the vortex core in an isotropic $s$--wave 
superconductor. Hess was the first who has experimentally confirmed 
the existence of core states or core excitations in ${\rm NbSe_2}$ 
superconductor\cite{hess}. These states manifest themselves in the local
density of states (LDOS) as a large peak near the Fermi surface. Similar
experiments carried out for HTSC have shown a rich structure of the 
vortex states. Namely, Maggio--Aprile {\it at al.} have found a splitting 
of the central peak in ${\rm YBa_2Cu_3O_{7-x}}$\cite{MA}. The resulting 
peaks are separated by about 11 meV. On the other hand the numerical 
solution of the Bogoliubov--de Gennes (BdG) equations for $d$--wave 
superconductivity (DSC)  indicates the absence of such a splitting.\cite{Wang} 
It was later explained by Franz and Tesanovi\v{c} that there are no 
truly bound states in $d$--wave superconductors and all the states are 
extended with continuous energy spectrum\cite{FT}. They tried to 
explain this discrepancy by introducing an additional $d_{xy}$ component
of the superconducting order 
parameter. The resulting $d_{x^2-y^2}+id_{xy}$ gap is nodeless and leads to
an exponential decay of the core states. Generally, the presence of the
bound states is expected for any nodeless gap. 
Although, $d_{x^2-y^2}$ is commonly accepted
as a dominant symmetry of the order parameter an additional 
component, that closes the nodes in the $x=\pm y$ directions, 
can be induced by the external magnetic field \cite{krila}.
However, results presented in Ref. \cite{kita} suggest 
that qualitative agreement with the STM 
experiments can be obtained also for a pure $d$--wave superconductivity, provided
that the magnetic field is strong enough. Moreover, the presence of the additional
component may not lead to an essential modification of LDOS.
The presence of the core states in BSCCO, unlike YBCO, is not well established.
Small peaks at $\pm 7$ meV have been reported in Ref. \cite{pan}. On the other
hand, the core states have not been observed in other STM
experiments \cite{renner}. Instead, a  gaplike structure has been
found in the center of the vortex core. It has been identified as
a pseudogap that evolves smoothly into the superconducting gap 
away from the vortex core. As the origin of the pseudogap is
still under debate, investigation of the vortex structure 
can shed a new light on this problem. 

The pseudogap has been investigated 
with the help of various experimental
techniques like: 
angle--resolved photoemission \cite{arpes1,arpes2,arpes3},
intrinsic tunneling spectroscopy \cite{krasnov1,krasnov2}, 
NMR \cite{nmr1,nmr2},
infrared \cite{infra} and transport \cite{transport} measurements.
Aside from other approaches a $d$--density--wave (DDW) 
state has recently been proposed to explain 
the pseudogap \cite{chak,nayak1,nayak2}. 
The DDW scenario has been investigated to verify whether
it actually applies to the pseudogap regime. For the details we refer to
considerations concerning  the transport 
properties \cite{tr1,tr2},  nonmagnetic \cite{imp1,imp2}
and magnetic  \cite{morr} impurities
as well as the phonon self--energy \cite{ph1,ph2} in the DDW phase. 
According to this hypothesis the pseudogap opens due to
condensation of electron--hole pairs with a nonzero angular
momentum ($l=2$). As a result there are staggered fluxes 
originating from the orbital currents, which alter from
one plaquette to the neighboring one. The DDW order
breaks the time reversal, rotational and translational
invariance, preserving combination of arbitrary 
two of them. There exists a similar approach that, however, 
does not break the translational symmetry \cite{varma}.

As the DDW and DSC orders compete \cite{chak,gosh}, one may expect
an enhancement (or appearance) of the DDW gap
near the vortex core, where the superconductivity is 
suppressed. It is possible due to the insensitivity of the DDW order 
to the magnetic field \cite{chakravarty}.
In the present paper
we investigate the vortex structure assuming the DDW scenario
of the pseudogap. Our approach explains the tunneling spectra 
obtained in the vicinity of the vortex core in BSCCO\cite{renner}. 
We analyze the vortex structure in a wide range of doping level and find
crucial differences between under-- and overdoped regimes. It was shown 
 in Ref. \cite{imp1} that depending on the 
carrier concentration one may expect pure DDW, mixed DDW+DSC, and
pure DSC ordering. Our results suggest the possibility of an
additional phase, where the DDW order occurs only inside the vortex 
cores. In the framework of the SU(2) slave--boson theory this possibility has
been investigated by Kishine {\it et al.}\cite{lee}. As the 
vortices are often pinned at impurities \cite{pan},  we  
investigate the electronic structure also in such a case. 
Finally, we discuss a possible coupling between the staggered currents
and the antiferromagnetic order, that has recently been 
observed in the vicinity of vortices.\cite{lake}  

\section{The model}

Our starting point is an effective Hamiltonian
that describes a system with coexisting DDW and DSC orders
\cite{imp1} 
\begin{equation}
H=H_{\rm kin}+H_{\rm DSC}+H_{\rm DDW} \label{H}.
\end{equation}
The kinetic part of the Hamiltonian is given by
\begin{equation}
H_{\rm kin} = -t\sum_{\langle ij\rangle \sigma} 
\eta_{ij}\left(\Av \right)\, c^{\dagger}_{i\sigma}c_{j\sigma} -
\mu \sum_{i\sigma} c^{\dagger}_{i\sigma}c_{i\sigma},  
\label{HKIN}
\end{equation}  
where $t$ is the nearest--neighbor hopping integral and $\mu$ is the
chemical potential.
 $c^\dagger_{i\sigma}$ ($c_{i\sigma}$)
creates (annihilates) an electron with spin $\sigma$  at site $i$. 
According to the Peierls substitution
\cite{peierls} the magnetic field enters the
kinetic part of the Hamiltonian
through a phase factor $\eta_{ij}\left(\Av \right)$:
in the presence of magnetic field 
the hopping between sites $i$ and $j$ is accompanied
by acquiring of an additional phase, given by
\begin{equation}
\eta_{ij}\left(\Av \right)=
\exp\left(\frac{ie}{\hbar c} \int^{\Rv_{i}}_{\Rv_{j}}
\Av\cdot d\lv\right).
\end{equation} 
$H_{\rm DSC}$ is the nearest--neighbor pairing
responsible for the $d$--wave superconductivity
\begin{equation}
H_{\rm DSC}=
\sum_{\langle ij\rangle}\left(c^{\dagger}_{i\uparrow}c^{\dagger}_{j\downarrow}
\Delta_{ij}
+c_{i\downarrow}c_{j\uparrow} \Delta^{*}_{ij} \right),
\label {HDSC}
\end{equation}
where 
\begin{equation}
\Delta_{ij}=-\frac{V_{\rm DSC}}{2}\langle c_{i\downarrow}c_{j\uparrow}
-c_{i\uparrow}c_{j\downarrow}\rangle
\end{equation}
 is the superconducting order
parameter.  
As we do not specify the mechanism responsible 
for pairing,  $V_{\rm DSC}$ is assumed to be
field--independent.\cite{nasz2} The DDW state occurs due to
\begin{equation}
H_{\rm DDW}=\sum_{\langle ij\rangle\sigma} 
(-1)^i W_{ij}\,\eta_{ij}\left(\Av \right) c^\dagger_{i\sigma}c_{j\sigma},  
\label{HDDW}
\end{equation}
where 
\begin{equation}
W_{ij}=(-1)^i\frac{V_{\rm DDW}}{2}
\langle \eta_{ij}\left(\Av \right)c^{\dagger}_{i\sigma}c_{j\sigma}
-\eta_{ji}\left(\Av \right)c^{\dagger}_{j\sigma}c_{i\sigma}\rangle 
\end{equation}
is the
DDW amplitude. Note, that in contradistinction to the superconducting
order parameter magnetic field explicitly enters $W_{ij}$ 
\cite{chakravarty}.

The mean--field Hamiltonian can be diagonalized with
the help of the Bogoliubov--de Gennes (BdG) equations.
Namely, we introduce a set of new fermionic operators
$\gamma^{\left( \dagger \right)}_{n \sigma}$:
\begin{eqnarray}
c_{i \uparrow}&=& \sum_{l} u_{il} \gamma_{l \uparrow}
-v^*_{il}\gamma^{\dagger}_{l \downarrow}, \nonumber \\
c_{i \downarrow}&=& \sum_{l} u_{il} \gamma_{l \downarrow}
+v^*_{il}\gamma^{\dagger}_{l \uparrow}, \nonumber
\end{eqnarray}
where
\begin{equation}
\sum_{j}\left(
\begin{array}{cc}
{\cal H}_{ij} & \Delta_{ij} \\
\Delta_{ij}^* & -{\cal H}_{ij}^* 
\end{array}
\right)
\left(
\begin{array}{c}
u_{jl} \\
v_{jl}
\end{array}
\right)
=E_l
\left(
\begin{array}{c}
u_{il} \\
v_{il}
\end{array}
\right).
\label{BdG}
\end{equation}
Here, the single particle Hamiltonian is given by
\begin{equation}
{\cal H}_{ij}=\left[-t\delta_{i+{\bf \delta},j}
+(-1)^{i}W_{ij}\right]\eta_{ij}\left(\Av \right)
-\mu\delta_{ij},
\end{equation}
and both the order parameters are determined self--consistently by:
\begin{eqnarray}
W_{ij}&=&(-1)^i\frac{iV_{\rm DDW}}{2}\sum_l{\rm Im}
\left[u_{il}u_{jl}^*\eta_{ij}\left(\Av \right) \right. \nonumber \\
&&+\left. v_{il}v_{jl}^*\eta_{ij}^*\left(\Av \right)\right]
\tanh\left(\frac{E_l}{2kT}\right),
\label{BDGDDW} \\ 
\Delta_{ij}&=&\frac{V_{\rm DSC}}{2}\sum_{l}\left(
u_{il}v_{jl}^*+v_{il}^*u_{jl}\right)
\tanh\left(\frac{E_l}{2kT}\right).
\label{BDGDSC}
\end{eqnarray} 
In the presence of magnetic field the phase factor $\eta$ 
explicitly appears  in the expression for the DDW
order parameter [Eq.(\ref{BDGDDW})]. It differs 
the BdG equations used in this paper from those obtained 
in Ref. \cite{imp1}.
       
In order to compare the numerical results with STM data we
have calculated the local density of states (LDOS)
\begin{equation}
\rho_i\left(\epsilon\right)=-\sum_{l}\left[
|u_{il}|^2f'\left(E_l-\epsilon\right)
+|v_{il}|^2f'\left(E_l+\epsilon\right)\right],
\label{LDOS} 
\end{equation}  
where $f'(\epsilon)$ is the derivative of the Fermi distribution 
function $f(\epsilon)=\left[\exp\left(\epsilon/kT\right)+1\right]^{-1}$. 
$\rho_i(\epsilon)$ is proportional to the local differential 
tunneling conductance that could be measured in STM experiments.
We have also calculated the local electron concentration
\begin{equation}
n_i=2\sum_{l}
|u_{il}|^2 f(E_l) +|v_{il}|^2 f(-E_l),
\label{ni}
\end{equation}   
that determines, e.g., the effective charge of the vortex.

\section{Results}

We have carried out calculations for $35 \times 35$ square 
lattice with one superconducting flux quantum piercing this area.       
We have  taken the nearest--neighbor hopping integral 
as the energy unit and assumed $V_{\rm DDW}=1.6$, $V_{\rm DSC}=1.4$.  
In the absence of magnetic field the phase diagram of HTSC can 
qualitatively be reproduced for such values of the interaction strengths
(we refer to Ref.\cite{imp1} for the details).

\subsection{Structure of vortex}

\begin{figure}
\epsfxsize=8cm
\centerline{\epsffile{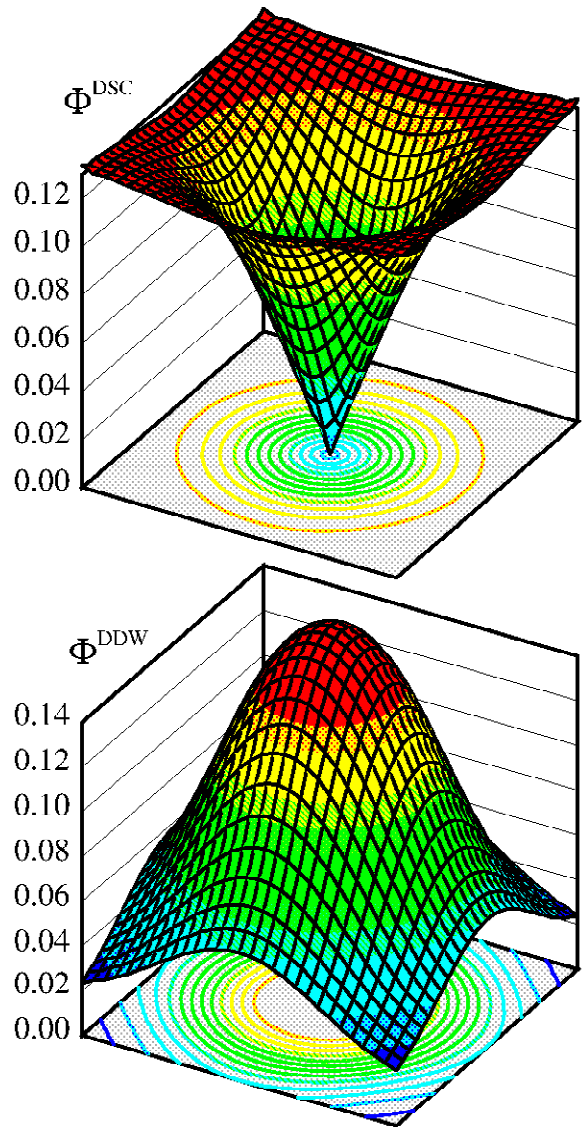}}
\caption{The vortex structure obtained for $\mu=-0.22$ ($\delta = 0.075$) 
and $kT=0.05$. The upper panel shows the DSC order parameter, whereas
the lower one corresponds to the DDW order.}
\label{cp022}
\end{figure}

\begin{figure}
\epsfxsize=8cm
\centerline{\epsffile{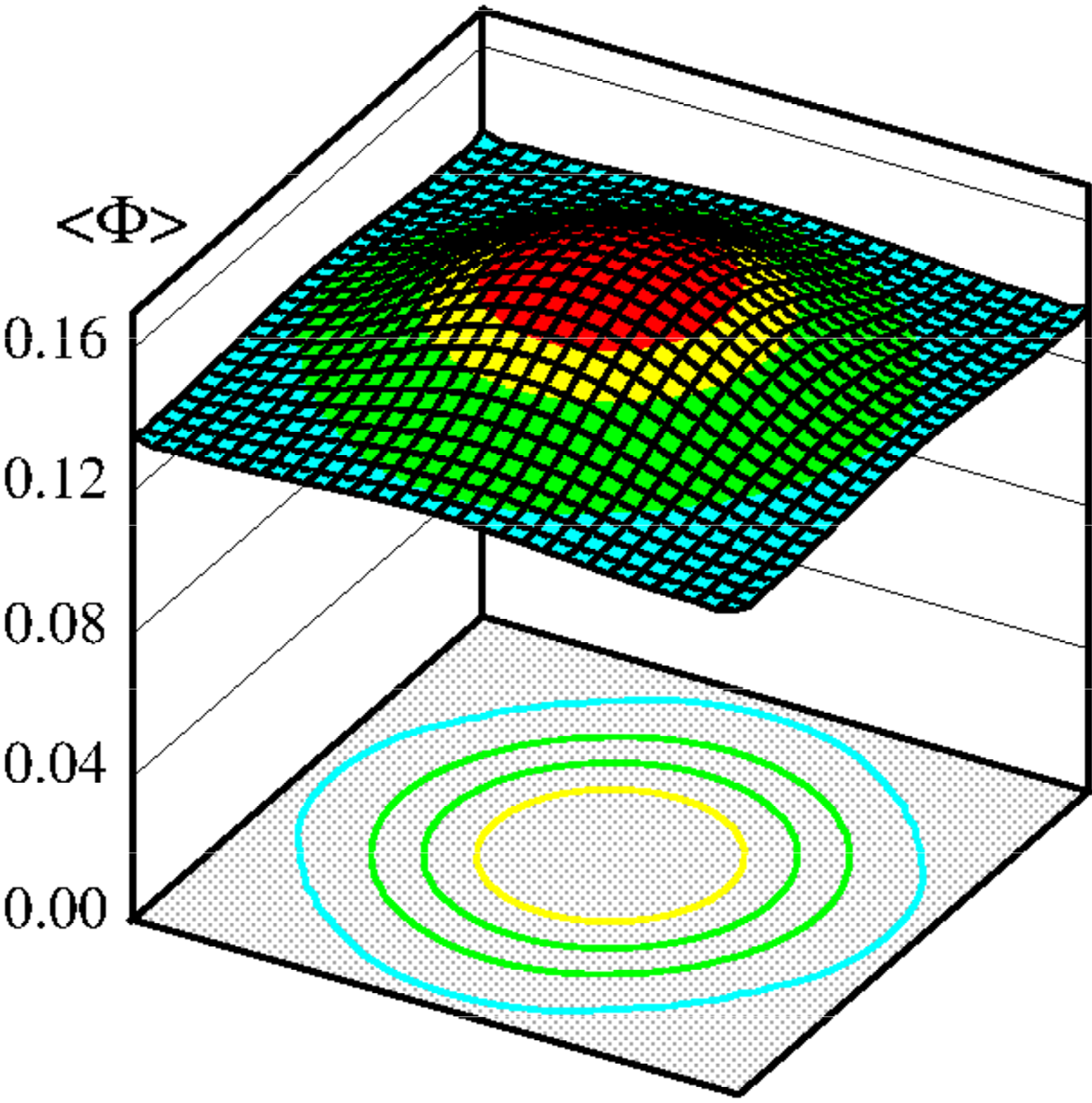}}
\caption{$<\Phi> \equiv \sqrt{\left(\Phi^{\rm DDW}_i\right)^2+
\left(\Phi^{\rm DSC}_i\right)^2 }$ calculated 
for the same parameters as in Fig. 
\ref{cp022}.}
\label{flat}
\end{figure}

We have analyzed the vortex structure for different values of the occupation 
number. For small concentration of holes, $\delta < 0.04$ ($\delta=1-n$), 
the DDW order dominates and superconductivity
is completely suppressed. Therefore, there is no vortex for such an occupation
number and the DDW state is homogeneous. For larger doping the
DDW and DSC orders coexist. For a doping slightly higher than
$\delta=0.04$  both the phases coexist in the whole sample, as
depicted in Fig. \ref{cp022}. 
The presented magnitudes of the order parameter
are defined by
\begin{eqnarray}
\Phi^{\rm DSC}_i &=&\frac{1}{4}\left(
\Delta_{i,i+{\bf\hat x}}+\Delta_{i,i-{\bf\hat x}}  
-\Delta_{i,i+{\bf\hat y}}-\Delta_{i,i-{\bf\hat y}}  
\right) \label{OPDSC} \\
\Phi^{\rm DDW}_i &=&\frac{1}{4}\left(
W_{i,i+{\bf\hat x}}+W_{i,i-{\bf\hat x}}  
-W_{i,i+{\bf\hat y}}-W_{i,i-{\bf\hat y}}  
\right) \label{OPDDW} 
\end{eqnarray}
\noindent
DDW and DSC orderings compete with each other \cite{chak}. 
Therefore, reduction of one of them
enhances the other one. As the DDW order is hardly affected by the magnetic field
it becomes strongly enhanced in the vortex core where the superconductivity is suppressed.
In order to prove that this mechanism is responsible for the enhancement of the
DDW order in the vortex core we have plotted 
$<\Phi> \equiv \sqrt{\left(\Phi^{\rm DDW}_i\right)^2+
\left(\Phi^{\rm DSC}_i\right)^2 }$ (see Fig. \ref{flat}). Note, that this quantity is almost
constant.  

\begin{figure}
\epsfxsize=8cm
\centerline{\epsffile{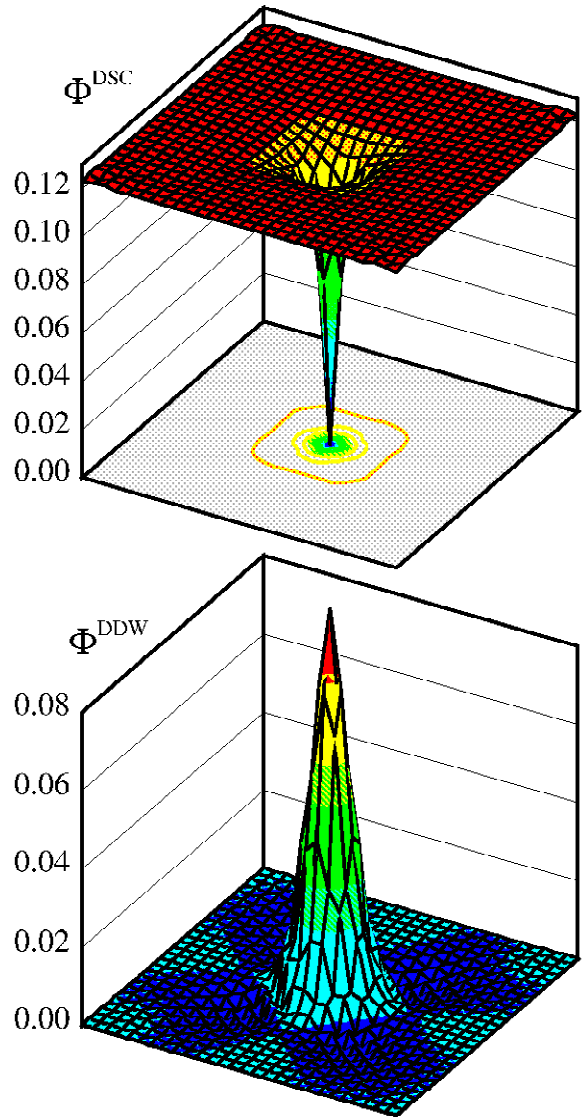}}
\caption{The same as in Fig. 1, but for $\mu=-0.38$ ($\delta=0.16$).}
\label{cp038}
\end{figure}

Farther increase of doping ($\delta > 0.1$) destroys the DDW order in a homogeneous
system.\cite{imp1} However, this ordering can be restored in the vortex core. 
Such a case is presented in Fig. \ref{cp038}. The local coexistence of DDW and DSC
takes place for the doping up to
$\delta \simeq 0.16$. Therefore, due to the suppression of superconductivity, 
magnetic field significantly enhances the
doping regime where the DDW order occurs. For higher doping the DDW order
is not restored in the vortex core. 
It originates from the fact that for sufficiently strong doping 
DDW does not exist even in the absence of superconductivity ($V_{\rm DSC}=0$).

\begin{figure}
\epsfxsize=8cm
\centerline{\epsffile{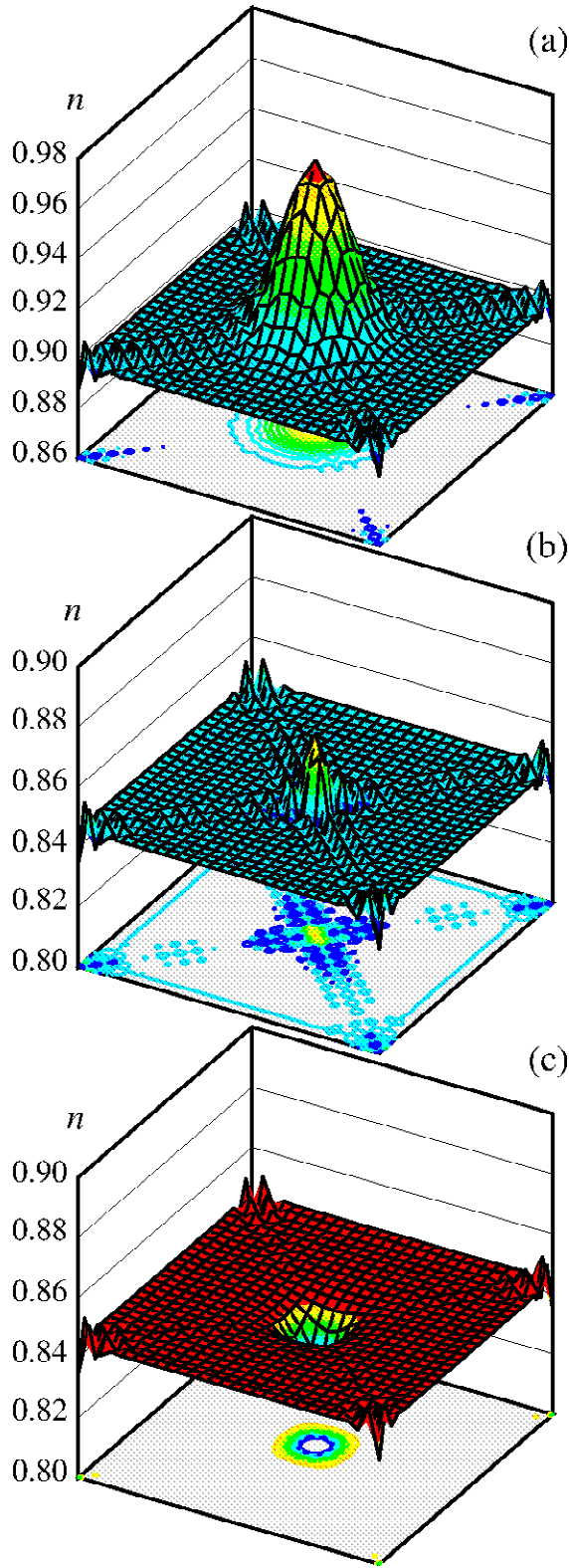}}
\caption{The electron concentrations in the vicinity of vortex.
 The upper panel (a) has been obtained for ($\delta=0.1$ and
$kT=0.05$. The middle (b) and lower (c) panels show the results for
$kT=0.05$ and $kT=0.1$, respectively. In cases (b) and (c) $\delta=0.16$. }
\label{charge}
\end{figure}

Another interesting feature, that originates from   
the presence of the DDW order is related to charge of the vortex.
For small doping the vortex charge is negative (i.e., the electron concentration
in the vortex core is higher than outside the vortex, see Fig. \ref{charge}a). 
As the doping increases the electron concentration in the vortex core decreases
and for $\delta > 0.16$ the vortex core becomes positively charged.
The inversion of the vortex charge occurs when the DDW order disappears.
It originates from the changes of LDOS: the DDW gap opens
in the center of the band and enhances the density of states below
the Fermi level. This inversion may also be induced by changing
of temperature, provided that the DSC transition temperature is larger
than those of the bare DDW state (overdoped region). Namely, at low temperature the 
DDW order is restored at the vortex core and, therefore, vortex is
negatively charged (see Fig. \ref{charge}b). At sufficiently high temperature
the DDW order does not occur and the vortex is positively charged 
(Fig. \ref{charge}c).
Experimental observation of this feature would certainly support the 
DDW scenario of the pseudogap.

\subsection{Local density of states}

\begin{figure}
\epsfxsize=8cm
\centerline{\epsffile{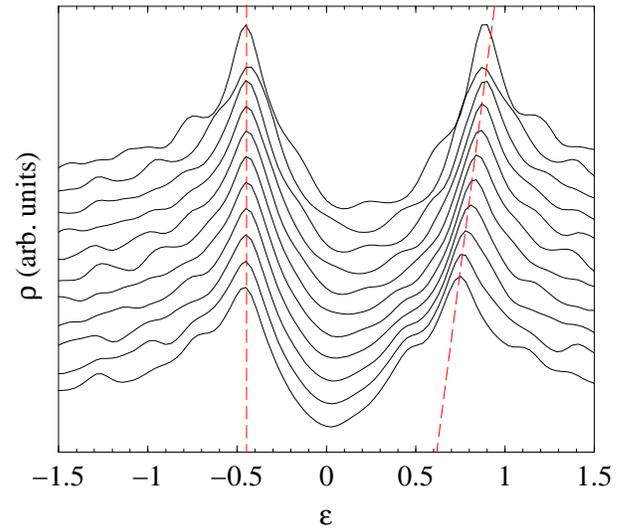}}
\caption{LDOS calculated for $\delta=0.075$ and $kT=0.05$ for
various distances from the vortex center. 
The uppermost curve corresponds to the vortex center, whereas the lowest
one to the distance of 10 lattice constants from the center.}
\label{ldos022}
\end{figure}

The first STM measurements of the vortex core in BSCCO did reveal
neither the existence of the core states nor the zero bias conductance
peak (ZBCP) expected for the pure DSC state \cite{renner}.
The presence of the core states in BSCCO has later been reported
in Ref. \cite{pan}. The discrepancy between both the experiments
remains, however, unexplained. In our approach
the DDW and DSC order parameters have nodes in the same parts of the Fermi surface
and the bound states do not occur 
(inclusion of components with other symmetries is needed).
However, we show that the DDW scenario explains the qualitative features
of LDOS reported in Ref. \cite{renner}.  In Fig. \ref{ldos022} we present
the LDOS for different distances  from the vortex center.   
For such occupation number both the DDW and DSC gaps are present
also away from the vortex. When approaching the vortex center the DSC gap,
that exists at the Fermi level,
vanishes and the DDW gap becomes strongly enhanced.
The DDW gap opens 
in the center of the band and, therefore, is responsible for the asymmetry of
the LDOS inside the vortex. In particular, the peak at positive bias
shifts outwards when approaching the center of vortex, whereas the peak
at negative bias does not move.
Such an asymmetry has recently been observed in BSCCO \cite{renner}. 
For larger doping this feature 
is even more pronounced since there is no DDW gap away from the vortex
and the Fermi level is much below the center of the band.	

\begin{figure}
\epsfxsize=8cm
\centerline{\epsffile{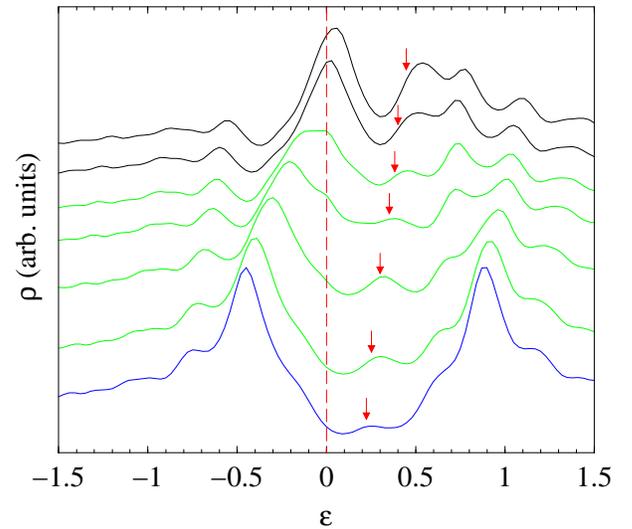}}
\caption{Doping dependence of LDOS calculated at the vortex center at
$kT=0.05$.
The curves from the bottom to the top correspond to the dopings:
$\delta=0.075,0.1,0.12,0.15,0.16,0.17,0.18$. The dashed line shows the Fermi
level, whereas the arrows indicate the center of band.}
\label{uo}
\end{figure}

The LDOS at the vortex center evolves smoothly with doping.
This is depicted in Fig. \ref{uo}.
The lowest curve corresponds to the case when DDW and DSC orders
coexist in the bulk material. The following four curves 
are obtained for DDW and DSC orders coexisting in the vortex, whereas the
two topmost curves are obtained for a  purely DSC vortex.  
For small
doping the DDW gap does not allow for the formation of the ZBCP. 
The DDW gap
decreases with increasing doping and  the
ZBCP gradually develops in the overdoped regime.  

\begin{figure}
\epsfxsize=8cm
\centerline{\epsffile{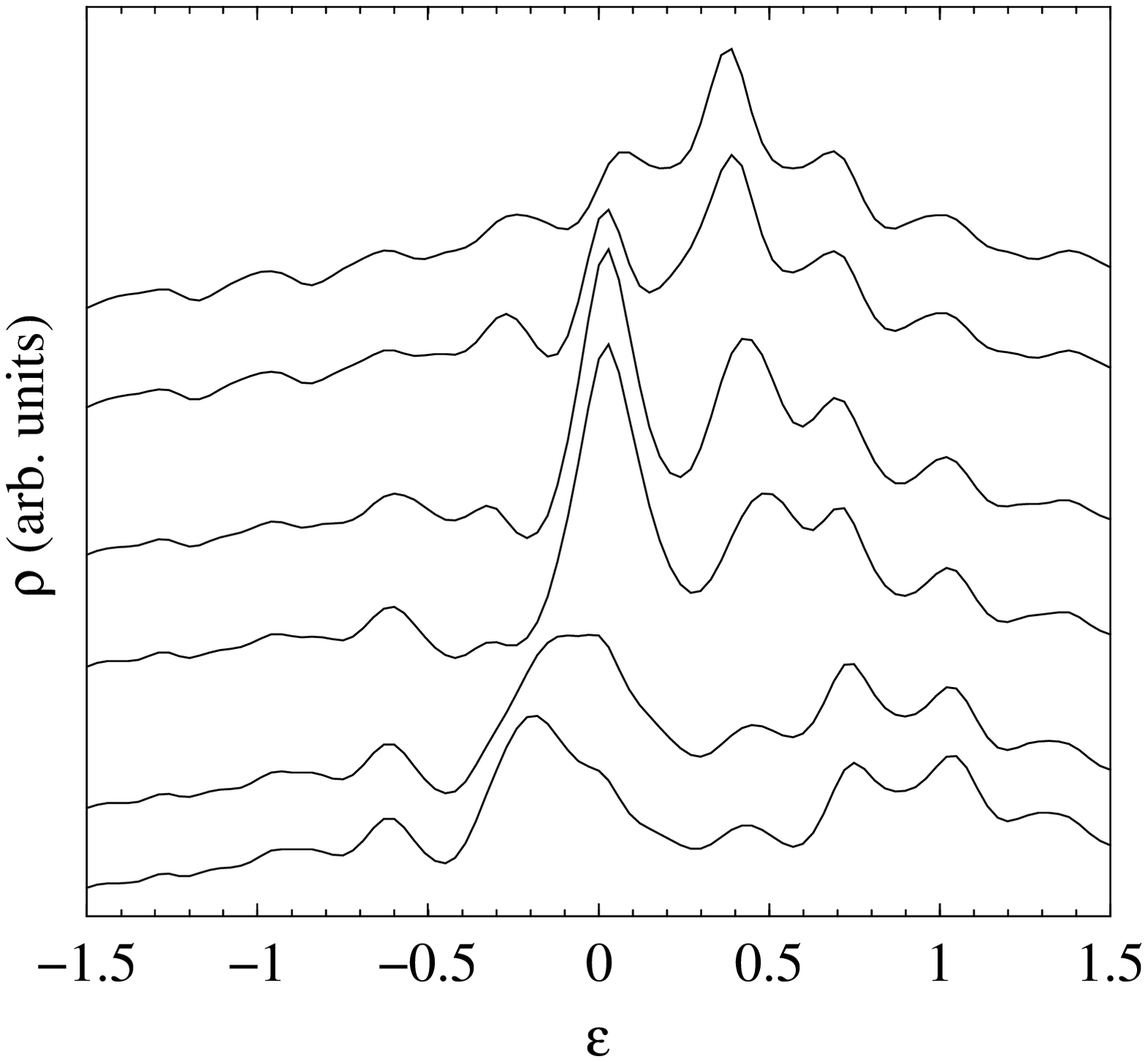}}
\caption{Temperature dependence of LDOS calculated at the vortex center for
$\delta=0.16$. The curves from the bottom to the top correspond to the 
temperatures:
$kT=0.01,0.05,0.10,0.15,0.20,0.25$. The 
broadening has been assumed to be temperature independent.}
\label{temp}
\end{figure}

In the overdoped regime a  similar evolution of LDOS, 
connected with the vanishing of the DDW gap,
can also be caused by the increase of temperature. 
Fig. \ref{temp} shows such a situation for $\delta=0.16$. 
At low temperature the DDW gap strongly suppresses the ZBCP, however
a remnant of this peak is visible at the Fermi level. At temperature
$kT=0.1$ the DDW gap closes and ZBCP becomes strongly enhanced.
Farther increase of temperature destroys superconductivity, ZBCP
disappears and the normal--state van Hove singularity appears 
in the center of the band (the uppermost curve, $kT=0.25$).

\begin{figure}
\epsfxsize=8cm
\centerline{\epsffile{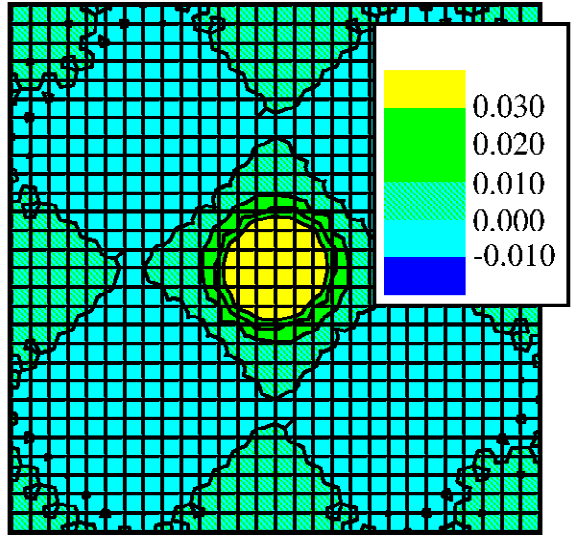}}
\caption{ 
The contour plot of the DDW order parameter in the vicinity
of the vortex calculated for $\delta=0.16$ and $kT=0.01$.}  
\label{checkerboard}
\end{figure}

Decay of the DDW order, when moving away from the vortex core,
is doping dependent. In the underdoped case the DDW order parameter
decreases monotonically from the maximum in the vortex core
to the value of the bulk DDW+DSC state. 
In the overdoped case the DDW order vanishes outside the vortex, however
the decay in not monotonic but rather oscillating. There exist lines
where the DDW order changes sign, what corresponds to reverted circulation
of the staggered currents (see Fig. \ref{checkerboard}). 
The resulting pattern posses a fourfold symmetry 
and has a checkerboard--like  modulation. However, the amplitude
decays rapidly and disappears at the distance of several lattice constants.
The size of the checkerboard squares is comparable to the
coherence length (a few lattice constants) and increases with decreasing 
doping. Similar pattern has recently been observed in the STM imaging
of slightly overdoped BSCCO \cite{hoffman}. The staggered flux alone 
may be to weak to explain the antiferromagnetic order observed 
in the vicinity of the vortex core \cite{lake}. However,
presence of DDW order may stabilize the antiferromagnetic 
ordering of spins. Here, two mechanisms can be taken into account:  
({\em i}) a coupling between the staggered magnetic field and spins,  
({\em ii}) changing of the local electron concentration  
towards half--filling, where the antiferromagnetism is 
stable (see Fig. \ref{charge}a).  
This tempting hypothesis should, however, be verified within
a separate study (e.g., one can extend the approach presented 
in Ref. \cite{kolor}). This problem is currently under investigation. 

\subsection{Vortex pinned at impurity}

As a large number of vortices is pinned at impurities \cite{pan}
we have also analyzed the vortex structure and LDOS
in such a case. Investigation of the LDOS in the
vicinity of a nonmagnetic impurity has recently
been proposed as an experimental test of the DDW scenario
\cite{imp1,imp2}. 
Since the DDW order is strongly 
enhanced in the vortex core one can expect a 
nontrivial modification of the vortex structure. In
order to account for the presence of a nonmagnetic
impurity we have extended the
Hamiltonian [Eq. (\ref{HKIN})] by
\begin{equation}
H_{\rm imp}= U \sum_{\sigma}
c^{\dagger}_{0 \sigma} c_{0 \sigma},   
\end{equation}
where $U$ is the strength of the impurity located at 
$\Rv_0$. Due to the presence of the impurity 
the single particle Hamiltonian in the BdG equations
acquires an additional term
\begin{equation}
{\cal H}_{ij} \rightarrow {\cal H}_{ij}
+U \delta_{i0} \delta_{j0}
\end{equation} 

\begin{figure}
\epsfxsize=8cm
\centerline{\epsffile{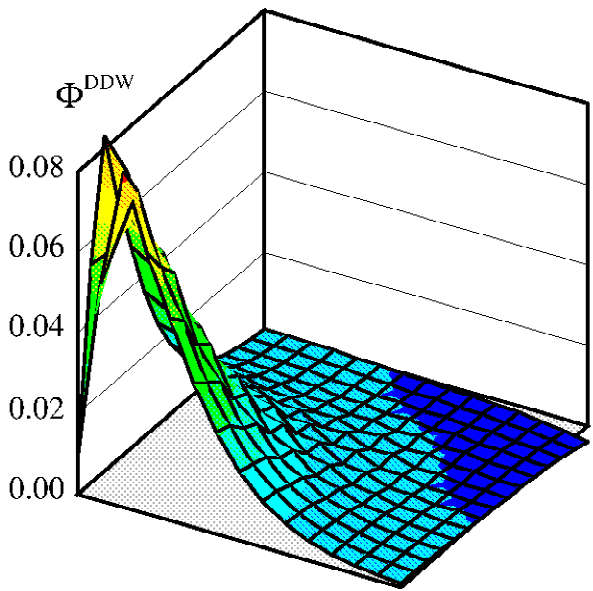}}
\caption{DDW order in the vicinity of vortex that is pinned at impurity.
We have used $\delta=0.1$, kT=0.05 and $U=2$. The vortex center is 
at the left corner of figure.}
\label{im1}
\end{figure}

\begin{figure}
%
\epsfxsize=8cm
\centerline{\epsffile{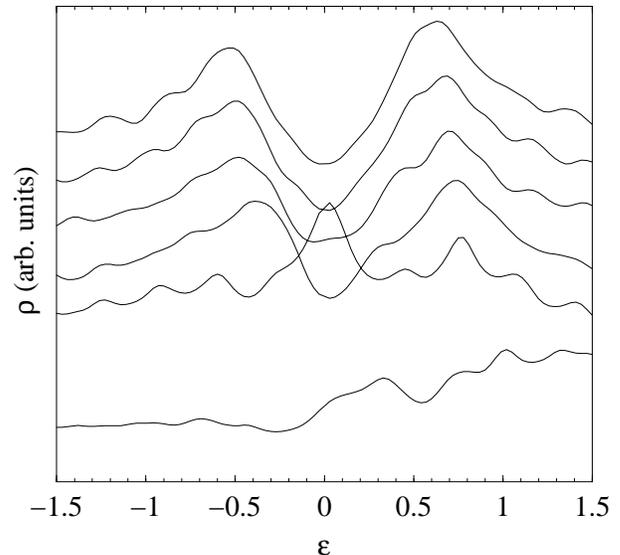}}
\caption{
LDOS calculated at various distances from the vortex center. 
The model parameters are the same as used in Fig 9. 
The lowest curve corresponds to the vortex center, whereas the uppermost
one to the distance of 5 lattice constants from the center.
}
\label{im2}
\end{figure}

Fig. \ref{im1} shows the DDW 
order parameter
around the vortex that is pinned at a weak impurity 
with $U=2$. The DSC order parameter is absent
in the vortex core and, therefore, weakly affected
by the impurity. Contrary to this, the DDW order is
is completely suppressed at the impurity site, 
provided that $U$ is large enough. Then, the DDW currents
are absent on the bonds, which link the impurity site
with its nearest neighbors. As a result 
the DDW order is suppressed also at the neighboring
sites and has a maximum on a ring surrounding the vortex center. 
The spatial structure of the pinned vortex is reflected in the
LDOS, (see Fig. \ref{im2}). At the impurity site 
the most of the spectral weight is transfered to larger 
energies. At the nearest neighbor site the DDW order is still suppressed by impurity, 
whereas DSC gap is negligible as we are in the vortex core. 
This allows for the formation of the ZBCP, as can be inferred from
the second curve in Fig. \ref{im2}. With increasing distance from the impurity
both the order parameters 
develop and the ZBCP disappears. Here, one can observe
the asymmetric gap, as discussed before for a clean superconductor.  
Then, the DDW order parameter decays and the pure DSC order parameter 
achieves its bulk value (the uppermost curve in Fig. \ref{im2}).

The above results show important qualitative differences between
the structure of vortices uninfluenced and influenced by impurities. 
Assuming the DDW origin of the pseudogap, 
the ZBCP should occur only in the latter case.   
Therefore, experimental verification of the DDW scenario  requires
the ability to independently map the vortex and impurity locations.
Such an experimental technique has recently been described in Ref.
\cite{pan}.
        
\section{Summary and remarks}

In conclusion, we have investigated the the vortex structure
in the system with coexisting DSC and DDW correlations.
Changing the carrier concentration we have found four phases:
({\em i}) pure DDW phase, ({\em ii}) DDW and DSC orders coexisting
in the whole system, ({\em iii}) DSC order existing in the bulk and
DDW order only in the vortex core, and ({\em iv}) pure
superconducting vortex. The system goes from the phase ({\em i})
to ({\em iv}) when the doping increases. The actual positions of the phase 
boundaries are model--dependent, and may change when the electron 
correlations are taken into account.
In the case ({\em ii}) the DDW order parameter is strongly enhanced
in the vicinity of the vortex core. It is due to the competition between
DSC and DDW orderings. The presence of the DDW gap in the vortex
[phases ({\em ii}) and ({\em iii})] does not allow for the formation of
ZBCP. In this case, the evolution of the DSC gap into DDW one, when 
approaching the vortex center, provides a natural explanation of the gap 
asymmetry reported in Ref. \cite{renner}. This asymmetry originates from
the fact that the DDW and DSC gaps open at different energies. 

As the DDW order suppresses
the ZBCP, this peak can be observed in the phase ({\em iv}), where 
the system is purely superconducting. However, one can expect the
occurrence of the ZBCP when the vortex is pinned at impurity. 
This is due to the absence of the staggered currents in the close 
neighborhood of the impurity.  

As DDW and DSC gapes have nodes in the same directions, 
the core bound states do not occur in our approach.
However, an additional component of the order parameter,
for which the gap is nodeless, may lead to the splitting
of the zero--bias peak in LDOS. 

The presented modification of the vortex structure originates 
predominantly from the competition between DSC and DDW orders.
A direct coupling of the DDW order 
to the magnetic field is of minor importance.
Therefore, one can argue that the presented results remain valid also
for other non--superconducting order parameter, that competes with DSC
and is weakly affected by the magnetic field. 

\acknowledgments

This work was supported in part by the Polish State Committee
for Scientific Research, Grant No. 2 P03B 050 23.

\end{document}